# An experimental search for an explanation of the difference between beam and bottle neutron lifetime measurements.

M. F. Blatnik[1,2], L. S. Blokland[3], N. Callahan[4], J. H. Choi[5], S. Clayton[1], C. B Cude-Woods[1], B. W. Filippone[2], W. R. Fox[3], E. Fries[2], P. Geltenbort[6], F. M. Gonzalez[7], L. Hayen[5], K. P. Hickerson[2], A. T. Holley[8], T. M. Ito[1], A. Komives[9], S Lin[1], Chen-Yu Liu[10], M. F. Makela[1], C. L. Morris[1,*], R. Musedinovic[5], C. M. O'Shaughnessy[1], R. W. Pattie Jr.[11], J. C. Ramsey[7], D. J. Salvat[3], A. Saunders[7], S. J. Seestrom[1], E. I. Sharapov[12], M. Singh[1], Z. Tang[1], W. F. Uhrich[1], J. Vanderwerp[3], P. Walstrom[1], Z. Wang[1], A. R. Young[1,5,13]

[1]*Los Alamos National Laboratory, Los Alamos, NM, USA, 87545, USA*
[2]*Kellogg Radiation Laboratory, California Institute of Technology, Pasadena, CA 91125, USA*
[3]*Department of Physics, Indiana University, Bloomington, IN, 47405, USA*
[4]*Argonne National Laboratory, Lemont, IL 60439, USA*
[5]*Department of Physics, North Carolina State University, Raleigh, NC 27695, USA*
[6]*Institut Laue-Langevin, CS 20156, 38042 Grenoble Cedex 9, France*
[7]*Oak Ridge National Laboratory, Oak Ridge, TN 37831, USA*
[8]*Tennessee Technological University, Cookeville, TN 38505, USA*
[9]*DePauw University, Greencastle, IN 46135, USA*
[10]*University of Illinois, Urbana, IL 61801, USA*
[11]*East Tennessee State University, Johnson City, TN 37614, USA*
[12]*Joint Institute for Nuclear Research, 141980 Dubna, Russia*
[13]*Triangle Universities Nuclear Laboratory, Durham, NC 27708, USA*

*Corresponding author: email:cmorris@lanl.gov

**Abstract** The past two decades have yielded several new measurements and reanalysis of older measurements of the neutron lifetime. These have led to a 4.4 standard deviation discrepancy between the most precise measurements of the neutron decay rate producing protons in cold neutron beams and the most precise lifetime measured in neutron storage experiments. Here we publish an analysis of the recently published UCNτ data[1] aimed a searching for an explanation of this difference using the model proposed by Koch and Hummel[2].

**Introduction**

Koch and Hummel[2] suggests a new solution to the neutron lifetime enigma[3]. The neutron lifetime enigma arises from the 4.4 standard deviation difference between the lifetime measured for bottled neutrons[1] and measurements of lifetime from a beam of cold neutrons[4]. Koch and Hummel point out the beam experiments measure the decay rate of neutrons very close in time to their source whereas the bottle measurement use neutrons ~1000 s after their production. They postulate the existence of an excited state of the neutron, n*, that has a longer $\beta$-decay lifetime than the ground state, n, and that a transition could occur between these two states by γ-ray emission with a decay time shorter than the holding time used for bottle lifetime measurements.

**Experiment**

The UCNτ experiment[1] stores ultra cold neutrons (UCN) produced by moderating spallation neutrons from 800 MeV proton interactions in a tungsten target at the Los Alamos UCN Facility[5, 6]. The UCN are transported to and stored in a magneto-gravitational trap[7] for various holding times and then counted using a $^{10}B$ coated scintillator screen[8] that is lowered into the trap and the end of the holding period. Before the holding period there is a 50 s cleaning period where quasi-trapped neutrons are removed by lowering a cleaner and the scintillator into the upper part of the trap. A spectrum from a sum of runs is shown in Figure 1. Over the course of this experiment the holding time was varied from 20 s to 4000s and data were taken with two different loading times, 150 s and 300 s. This provides a wide range of times from the neutron birth until counting to test the Koch and Hummel hypothesis[2].

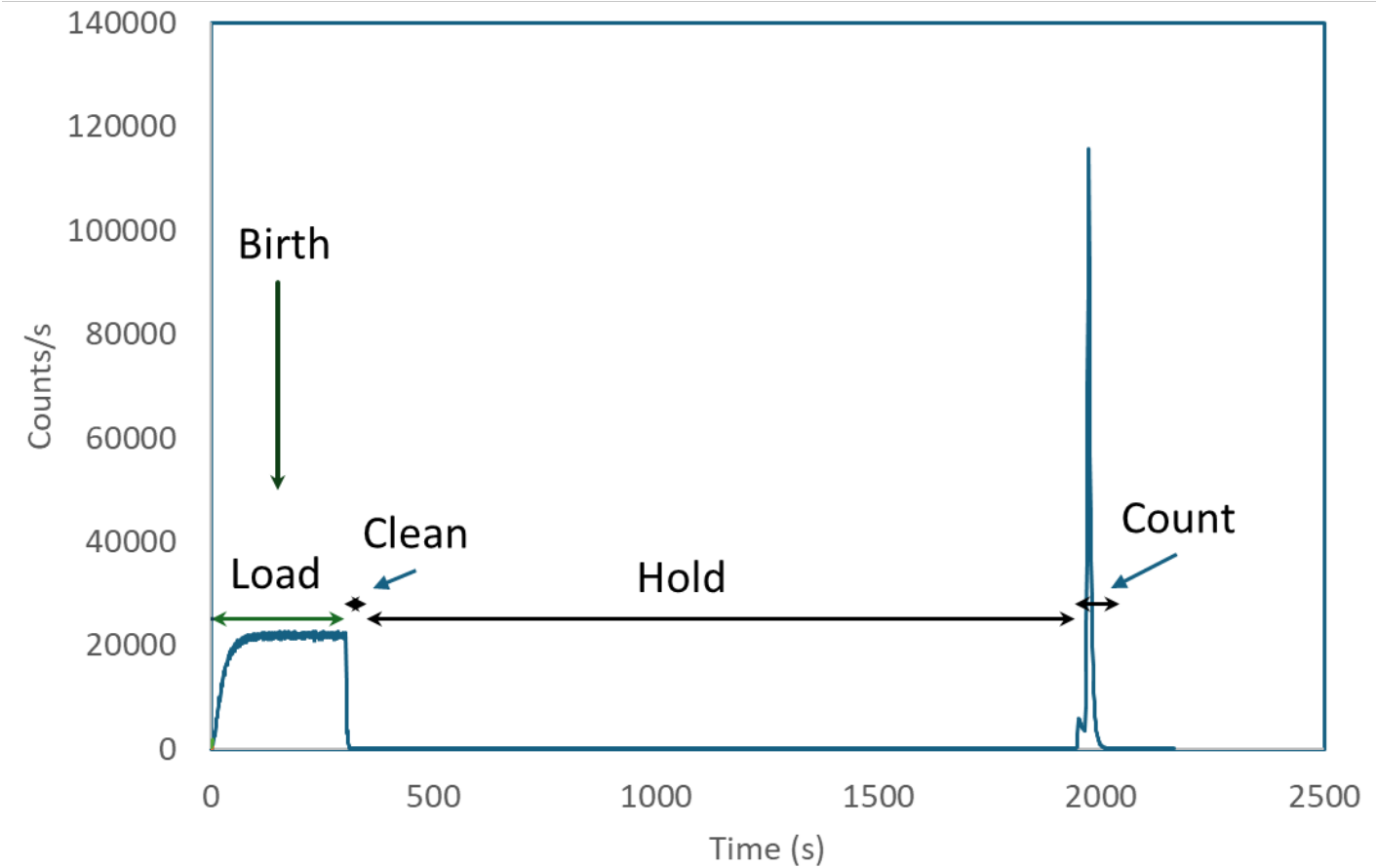

Figure 1) Summed spectrum from the UCN detector for a set of 1550 s holding time runs showing the counting sequence. The average time of birth is taken as the middle of the beam on time (Load).

**Analysis**

This model is described by four parameters, $f$, $\tau_b$, $\tau_\gamma$, and $\tau_t$, the fraction of neutrons born in the excited state, the beam lifetime, the decay lifetime of excited to ground state neutrons and the trap lifetime respectively. The neutron production methods and spectra are very different between the reactor source used for the beam experiments and in the accelerator driven spallation source used in the current experiment. Here we have varied the fraction of $n^*$, $f$, at $t = 0$ over a wide range but have constrained them to be equal, independent of production mechanism.

The coupled differential equations:

$$x(0) = f$$
$$y(0) = (1-f)$$
$$\frac{dx}{dt} = \left[ -\frac{x(t)}{\tau_\gamma} - \frac{x(t)}{\tau_b} \right]$$
$$\frac{dy}{dt} = \frac{x(t)}{\tau_\gamma} - \frac{y(t)}{\tau_t}$$

which give the time derivatives of the number of the excited neutrons, $x(t)$, and ground state neutrons, $y(t)$, in the trap versus time, $t$, in terms of the three decay times The solution to these equations is:

$$x(t) = f e^{-\frac{t(\tau_b + \tau_\lambda)}{\tau_b \tau_\lambda}}$$

$$y(t) = f \frac{-\tau_b \tau_t}{-\tau_b \tau_\gamma + \tau_b \tau_t + \tau_\gamma \tau_t} \left( e^{-\frac{t(\tau_b + \tau_\gamma)}{\tau_b \tau_\gamma}} - e^{\frac{-t}{\tau_t}} \right) + (1-f) e^{-\frac{t}{\tau_t}}.$$

The number of remaining neutrons, given by $x(t) + y(t)$, can be fitted to the bottle lifetime measurements of Gonzalez[1] with $\tau_b$ adjusted so that the decay rate at $t=0$ matches the beam lifetime measurement.

Since the data cover a wide range of times when surviving neutrons in the trap are counted (Figure 1) the result of lifetime fits as a function of $t$ should be sensitive to $\tau_\gamma$.

LA-UR-24-25619

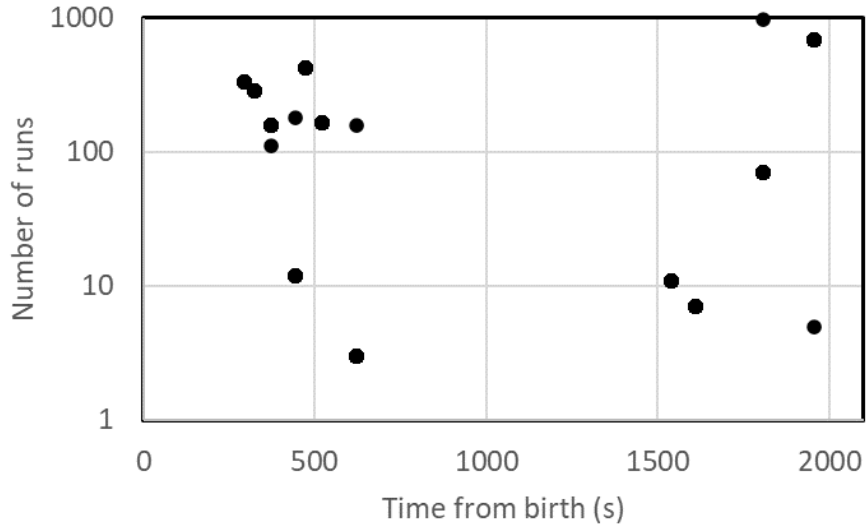

*Figure 2) Histogram of the runs as a function of time from the start of beam.*

Fits were performed by using the Yields, $Y_{0k}$, and uncertainties, $\Delta Y_{0k}$, that had been calculated for each run, k, in [1]

$$Y_{0k} = \frac{S_k - B_k}{N_k}$$

$$\Delta Y_{0k} = \frac{\sqrt{S_k + B_k}}{N_k}.$$

Where $S_k$ are the total counts in the foreground period, $B_k$ are the background counts measured after the foreground counting period, and $N_k$ are the counts in a monitor counter.

Several small corrections were the same as those in [1]. For each run, $k$, the expected yield, $Y_{calc,k}$, was calculated with a sliding normalization factor, $CF$, the eliminates sensitivity to slow time variations in the source output.

$$Y_{calc,k} = f \frac{-\tau_b \tau_t}{-\tau_b \tau_\gamma + \tau_b \tau_t + \tau_\gamma \tau_t} \left( e^{-\frac{t(\tau_b + \tau_\gamma)}{\tau_b \tau_\gamma}} - e^{\frac{-t}{\tau_t}} \right) + (1-f) e^{-\frac{t}{\tau_t}}$$

$$R_k = \frac{Y_k}{Y_{calc,k}}$$

$$CF = \overline{R_k}$$

This is necessary because the monitor counts neutrons in a different spectral region from those loaded into the trap, and the aging of the source changes the output spectrum[9]. Lifetime fits were performed for several different values of $\tau_\gamma$ by varying $\tau_t$ to minimize the chi square, $X^2$:

$$X^2 = \sum_k \frac{\left(Y_k - Y_{calc,k} CF_k\right)^2}{\left(\Delta Y_k\right)^2}.$$

Corrections for the statistical bias, gas pressure, and phase space evolution were applied as in Gonzalez[1].

Scan of $X^2$ as a function of $\tau_\gamma$ are plotted in Figure 3 for different values of $f$. For these the n* beta decay lifetime, was obtained from:

$$\frac{1}{\tau_{NIST}} = \frac{f}{\tau_t} + \frac{1-f}{\tau_b},$$

where $\tau_{NIST}$ is the lifetime measured in the beam experiment[4] of 887.7±2.2. This assume that the beam experiment counts n* and n in the beam with the same efficiency and that the decay proton from n* is not lost out of the detector or trap if the excitation energy is too large. We also assume that the UCN detector counts n and n* with the same efficiency.

These fits show no evidence for an excited neutron with a $\gamma$-ray decay lifetime longer than 139 s at a 95% confidence level for all possible n* fractions under these assumptions.

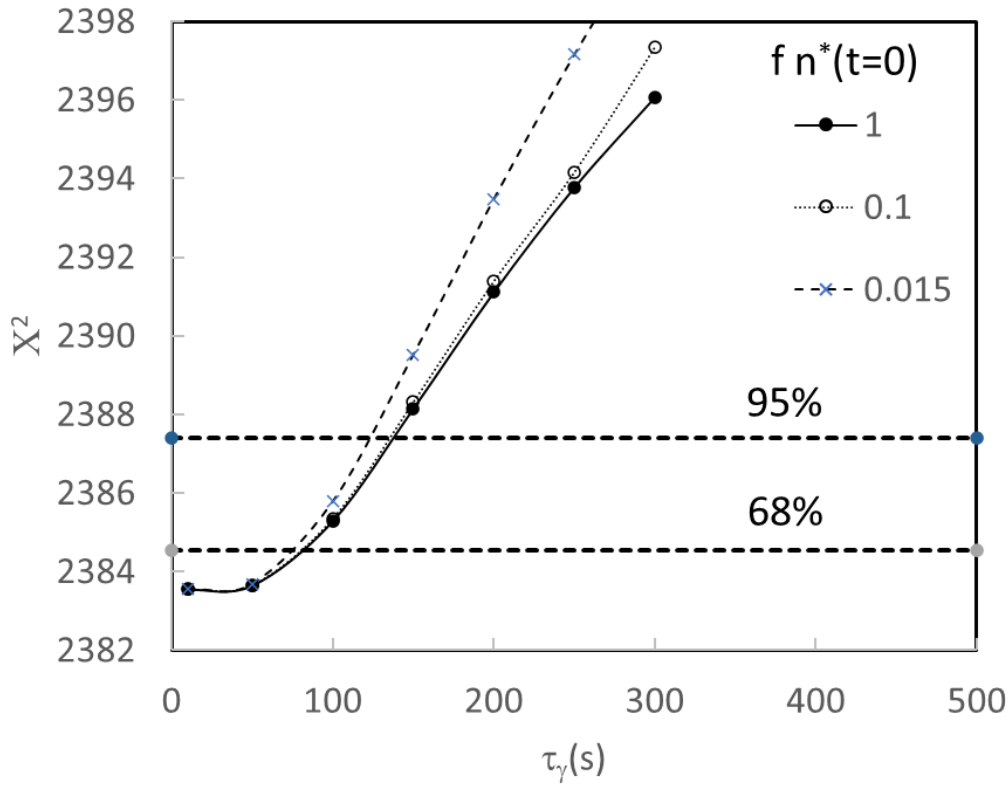

Figure 3) $X^2$ as a function of $\tau_\lambda$ for a range of starting n* fractions. The 68% and 95% confidence values are plotted as horizontal dashed lines.

The space available for the excitation energy of n* and the γ-ray decay lifetime are shown in Figure 4, reproduced from [2]. The blue and red boxes are regions excluded by previous experiments. The horizontal orange lines are upper bounds on the excitation energy under different assumptions discussed in [2]. The shaded areas at the left and right of the diagram show the time regions occupied by the beam and bottle experiments respectively. The shaded green area is excluded at the 95% confidence level by the current experiment.

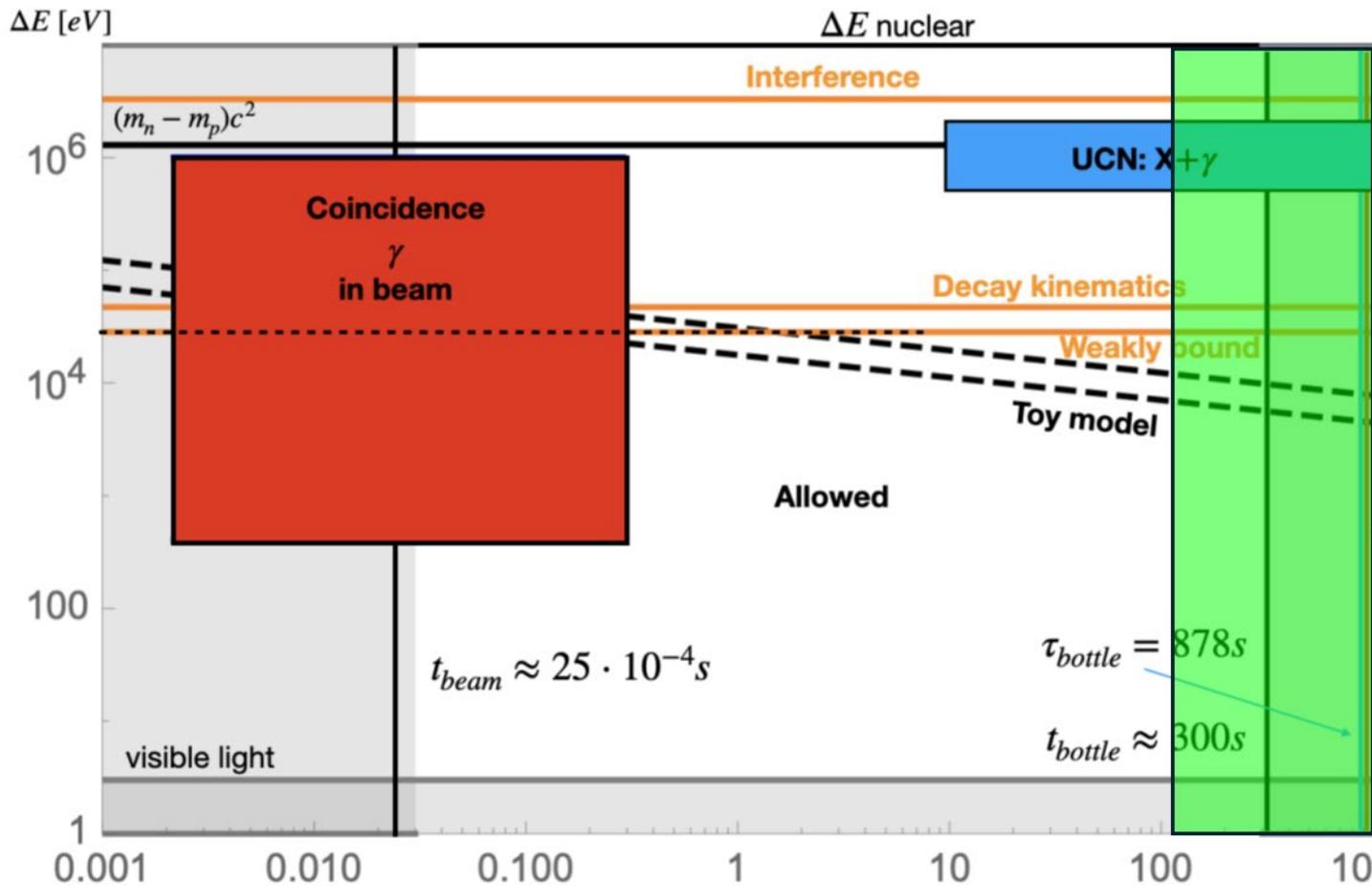

Figure 4) figure reproduced from Koch and Hummel[2] shown the model space and two regions that are excluded by previous experiments. The new 95% exclusion region form this work is shaded in green.

**Conclusion**

The data from a previous lifetime experiment[1] have been reanalyzed to test a new hypothesis from Koch and Hommel[2] for explaining the neutron lifetime enigma[3]. The data were refitted with a model that assumes an excited neutron that decays to the ground state with a lifetime shorter than the neutron lifetime, and that the two neutron states have different lifetimes. If the lifetime at birth is longer than the ground state the two lifetimes can be adjusted to agree with the decay rate measured in the beam experiment, and the lifetime measured in the storage experiment. This reanalysis shows no evidence for this model with a γ-ray decay lifetime longer than 139 s at the 95% confidence level for all fractions of n* at t=0 s .

**Acknowledgments**

This work is supported by the LANL LDRD program; the U.S. Department of Energy, Office of Science, Office of Nuclear Physics under Awards No. DE-FG02-ER41042, No. DE-AC52-06NA25396, No. DE-AC05-00OR2272, and No. 89233218CNA000001 under proposal LANLEDM; NSF Grants No. 1614545, No. 1914133, No. 1506459, No. 1553861, No. 1812340, No. 1714461, No. 2110898, No. 1913789, and No. 2209521; and NIST precision measurements grant.